# Interface induced high temperature superconductivity in single unit-cell FeSe on SrTiO$_3$(110)

Guanyu Zhou[1,a)], Ding Zhang[1,a)], Chong Liu[1,a)], Chenjia Tang[1], Xiaoxiao Wang[1], Zheng Li[1], Canli Song[1,2], Shuaihua Ji[1,2], Ke He[1,2], Lili Wang[1,2,b)], Xucun Ma[1,2] and Qi-Kun Xue[1,2,b)]

[1] *State Key Laboratory of Low-Dimensional Quantum Physics, Department of Physics, Tsinghua University, Beijing 100084, China*

[2] *Collaborative Innovation Center of Quantum Matter, Beijing 100084, China*

a)G.Y. Zhou, D. Zhang and C. Liu contributed equally to this work.

b)Authors to whom correspondence should be addressed. Electronic addresses: liliwang@mail.tsinghua.edu.cn and qkxue@mail.tsinghua.edu.cn

ABSTRACT

We report high temperature superconductivity in one unit-cell (1-UC) FeSe films grown on SrTiO$_3$ (STO) (110) substrate by molecular beam epitaxy. By *in-situ* scanning tunneling microscopy measurement, we observe a superconducting gap as large as 17 meV on the 1-UC FeSe films. Transport measurements on 1-UC FeSe/STO(110) capped with FeTe layers reveal superconductivity with an onset transition temperature ($T_C$) of 31.6 K and an upper critical magnetic field of 30.2 T. We also find that $T_C$ can be further increased by external electric field although the effect is weaker than that on STO(001) substrate.

The single unit cell (UC) films of FeSe on perovskite $SrTiO_3$(STO)(001) substrate (referred as FeSe/STO(001) hereafter) exhibits probably the highest superconducting transition temperature $T_C$[1-8] among all heterostructure systems discovered so far. A superconducting gap of 20 meV[1], almost one order of magnitude larger than that of bulk FeSe[9], was observed by scanning tunneling microscopy (STM). Angle resolved photoemission spectroscopy (ARPES) revealed a similar gap that closed at ~65 K[2-4], while transport measurement on this system covered by FeTe capping layers demonstrated a superconducting transition temperature $T_C$ > 40 K[5]. *Ex-situ* mutual inductance and *in-situ* four-probe measurements even indicated a transition temperature of 65 K[8] and 100 K[7], respectively, which were higher than the record $T_C$ = 55 K of bulk iron-based superconductors[10].

The advent of enhanced superconductivity in FeSe/STO(001) has instigated great interests in other interfacial systems[1,11-14]. For example, by replacing FeSe with $FeTe_{1-x}Se_x$ a superconducting gap of 17 meV was observed[14], while using $BaTiO_3$ and strained STO substrates led to a gap closing temperature up to 75 K[11,12]. The investigations reveal the crucial roles of interface charge transfer[2-4,6,15-20] and electron-phonon coupling[1,6,18-21]. In addition, stress caused by lattice mismatch between FeSe and STO substrate is also thought to be responsible for $T_C$ enhancement[4]. To figure out the key role of substrate, STO(110) substrate is of great interest because it resembles STO(001) in high density subsurface oxygen vacancies[22-24] but distinguishes itself by a rectangular in-plane lattice[25,26] and the correspondingly different dielectric constants[27,28]. Here, we investigated molecular beam epitaxy (MBE) growth of 1-UC FeSe films on STO(110) substrates (referred as FeSe/STO(110) hereafter) and studied the superconducting properties by combined *in-situ* STS and *ex-situ* transport measurement.

The growth of single unit-cell FeSe films on Nb-STO and insulating (110) substrates was carried out in ultra-high vacuum (UHV) STM-MBE systems. The STO(110) substrate has a nominal thickness of 0.50 ± 0.05 mm. To prepare clean surfaces, the Nb-STO(110) substrates were heated to 1100 ℃ for 30 minutes in UHV chamber, while the insulating STO(110) substrates were first pretreated by chemical etching in a 10%-HCl solution followed by thermal annealing in a tube furnace at 1080 ℃ for 3 hours, and then degassed at 600 ℃ for 30 minutes in UHV chamber. Stoichiometric FeSe films were obtained by co-evaporating Fe (99.995%) and Se (99.9999%) with a flux ratio of ~1:10 at a substrate temperature of 400 ℃ and post-annealing at 500 ℃ for several hours. Crystalline FeTe capping layers were obtained by co-evaporating Fe (99.995%) and Te (99.9999%) with a flux ratio of ~1:4 at a substrate temperature of 300 ℃. The growth rate was determined by Fe flux and approximately 0.2 UC/min for both FeSe and FeTe. In all STM/STS measurements, a polycrystalline Pt-Ir tip was used and the bias voltage was applied to sample. The STS was acquired at 4.6 K by using the lock-in technique with a bias modulation of 0.4 mV at 837 Hz. For *ex-situ* transport measurement, the resistance was measured by the standard lock-in technique (Stanford Research 830, $I_{ac}$=1 μA, 13 Hz). DC current-voltage characteristics were obtained by driving the current from the source to the drain and measuring the four-terminal longitudinal voltage with a pre-amplifier (Keithley 2400 as the current source, Stanford Research 560 as the amplifier). Cryogenic temperatures and high magnetic fields were realized in a commercial physical property measurement system (PPMS, Quantum Design).

One UC FeSe in β-phase consists of two Se layers sandwiching a Fe layer with an in-plane lattice constant of 3.78 Å and an out-of-plane lattice constant of 5.50 Å[29]. STO, the archetypical perovskite oxide, has a lattice constant of 3.91 Å[25]. Along the [110] direction, it is composed of alternating $SrTiO^{4+}$ and $O_2^{4-}$ layers and is hence polar. For charge compensation, the top surface transforms into a tetrahedrally coordinated titania layer reconstructed with $n \times 1$ periodicities (n=3, 4, 5, 6 corresponds to the reconstruction along the [001] direction, which are nearly energy-degenerate)[23,25,26]. Indeed, we evenly observed $4 \times 1$, $5 \times 1$ and $6 \times 1$ reconstructions on the Nb-STO(110) surface. Fig. 1(a) presents the morphology of a Nb-STO(110) substrate with a uniform 4×1 reconstruction.

Displayed in Fig.1(b)-(c) and (e)-(h) are the morphologies of 1-UC FeSe films on Nb-STO and insulating STO substrates, respectively. The 1-UC FeSe films on Nb-STO exhibit irregular bumps (Fig. 1b) and strong local distortion with a standard deviation of 1 Å (Fig. 1(c)). The Fourier transformation of Fig. 1(b) gives nearly the same lattice constants of a=b=3.9 Å along the two orthogonal directions. Despite such a strong local distortion, tunneling spectroscopy reveals a gap of 17 meV for 1-UC FeSe films on Nb-STO at 4.6 K (Fig. 1(d)). The gap is defined as half of the distance between the two coherence peaks. Such a large superconducting gap is similar in magnitude to most of the reported values in interface enhanced superconducting FeSe monolayers[2-4,30] but smaller than the maximum value of 20 meV[1,31].

In the case of 1-UC FeSe films on insulating STO, the morphology of the FeSe films follows the terrace-step structure of STO(110) substrate, as evidenced by a step height of 2.7 Å (Fig. 1(e)) which is equal to the out-of-plane lattice constant of STO along [110] direction[25], indicating a perfect layer-by-layer growth. The large-scale uniform films guarantee transport measurements shown later. In contrast to irregular bumps observed in Fig. 1(b), the 1-UC FeSe films on insulating STO exhibit nearly ordered stripes with periods consistent with the 6 ×1 and 5 ×1 reconstructions of STO(110) surface, i.e. ~ 2.4 nm for Fig. 1(f) and ~ 2.0 nm for Fig. 1(g). Moreover, the lattice constants are distinctly anisotropic. The Fourier transformation in Fig. 1(g) yields lattice constants of 3.9 Å and 3.7 Å along the [001] and [1-10] directions of STO, respectively. These values indicate that the lattice is statistically 3% expanded and 2% contracted in the two orthogonal directions, if compared with the isotropic lattice constants of 3.78 Å for bulk FeSe. The local distortion of FeSe films on insulating STO is weaker than that of FeSe films on Nb-STO, as evidenced from the comparison of standard deviation, i.e. 0.5 Å in Fig. 1(h) vs. 1 Å in Fig. 1(c), which could be the reason that the lattice anisotropy is resolved in the former case. Based on the above observation, we propose a model to show the epitaxial relation between FeSe films and STO(110) substrate. As depicted in Fig. 1(i)(j), three unit cells of FeSe coincide with two unit cells of STO along the [1-10] direction, whereas an one-to-one correspondence between FeSe and STO exists along the [001] direction. It is worth noting that either irregular bumps (Fig. 1(b)) or periodic stripes (Fig. 1(f) and (g)) could originate from vertical shift of FeSe films due to the FeSe strain resulting from lattice mismatch. We speculate that the contrast between bumps and stripes may correlate with different surface structure due to different pretreatment of Nb-STO and insulating STO substrates.

To establish that the gap opening as observed in Fig. 1(d) corresponds to a superconducting transition, we performed *ex-situ* transport measurements on the samples prepared on insulating STO substrates. The 1-UC FeSe films are capped by 10-UC FeTe layers which are insulating. This capping method has been successfully used to protect FeSe/STO(001) from ambient contamination/oxidation[5]. Figure 2(a) displays the sample resistance as a function of temperature at zero field (a schematic setup for transport measurements is shown in the inset). The resistance exhibits a linear dependence on temperature below 100 K, deviates from linearity at 47 K, and drops by three orders of magnitude at 16 K. By extrapolating the normal resistance and the superconducting transition curves, we obtain an onset temperature of $T_{onset}$ ~31.6 K, which is almost four times of the $T_C$ ~8 K for bulk FeSe[29]. The same transition is seen from another pair of contacts on the edge of the sample (right panel of Fig. 2(a)), attesting to the macroscopic homogeneity of our films. Our following data are from the central pair of contacts; the other pairs give essentially the same results. The inset in Fig. 2(b) shows the resistance as a function of temperature at various magnetic fields applied perpendicular to the film. We can clearly see that as the magnetic field increases, the superconductivity transition region becomes broader and $T_C$ shifts to lower temperatures. For a 2D superconductor, its response to an external magnetic field normal to the 2D plane is expected to follow the linear Ginzburg-Landau formula[24]:

$$\mu_0 H_{c2} = \frac{\Phi_0}{2\pi\xi^2}(1 - \frac{T_c}{T_{c,H_{c2}(0)}}) \quad (1),$$

here $\mu_0 H_{C2}$ is the upper critical field, $\xi$ the in-plane coherent length, and $\Phi_0 = h/2e$ the flux quantum. $T_{c,H_{c2}(0)}$ and $T_C$ represent the superconducting transition temperatures at zero field and at a finite field, respectively. Due to the 2D nature of this system, the superconducting transition spreads in a broad temperature window of more than 10 K. Here, we use the mid-point temperature $T_{mid}$ at which the resistance drops to 50% of the normal state resistance, namely $R(T_{mid}) = 50\% R_{norm}$, to estimate the upper critical field. As shown in Fig. 2(b), $T_{mid}$ indeed behaves linearly with the magnetic field up to 9 T, consistent with the formula (1). The linear fitting gives an upper critical field of $\mu_0 H_{c2(110)}$ ~30.2 T. The coherent length calculated by using Eq. (1) is $\xi_{(110)} = 3.3$ nm. These values are much larger than the thickness of a single UC FeSe (0.55nm), attesting to its 2D superconductivity nature. In the case of FeSe/STO(001)[5], the upper critical field is 56.8 T and the coherent length $\xi_{(001)} = 2.4$ nm (Fig. 2(b)). According to the BCS theory[32], the relation $\xi_{(110)} > \xi_{(001)}$ implies that the superconducting gap in FeSe/STO(110) system should be smaller than that in FeSe/STO(001).

It is consistent with the results that $\Delta_{(110)}$ ~17 meV < $\Delta_{(001)}$ ~20 meV[1] and $T_{onset\,(110)}$ ~31.6 K < $T_{onset\,(001)}$ ~40.2 K[5].

The 2D superconductivity is evidenced by the signature of Berezinski–Kosterlitz–Thouless (BKT) transition[33]. The voltage versus current $V(I)$ characteristics shown in Fig. 3(a) was measured at temperatures ranging from 16 to 45 K at zero magnetic field. The $V(I)$ curves exhibit a $V \sim I^\alpha$ power-law dependence, and the exponent changes systematically as expected for the BKT transition[33,34]. A detailed evolution of the α-exponent as a function of temperature is summarized in Fig. 3(b). With decreasing temperature, the exponent α deviates from 1 and approaches 3 at 24 K-the temperature identified as $T_{BKT}$. Additionally, the observed $R(T)$ characteristics are consistent with a BKT transition. Just above $T_{BKT}$, the resistance has the following temperature dependence: $R(T)= R_0 exp[-b(T/T_{BKT}-1)^{-1/2}]$, where $R_0$ and $b$ are material parameters[35]. Plotting $[dln(R)/dT]^{-2/3}$ against $T$ would therefore reveal a linear dependence with the extrapolation of $[dln(R)/dT]^{-2/3}$ to zero at $T_{BKT}$. As shown in Fig. 3(c), such an extrapolation yields $T_{BKT}$ = 24.1 K, which is in agreement with the result of the $V$-$I^\alpha$ analysis.

Similar to FeSe/STO(001)[5], FeSe/STO(110) is a superconductor dominated by n-type carriers. The Hall coefficient exhibits the same evolution with decreasing temperature, *i.e.* changing from positive to negative values at a temperature around 120-130 K before the superconducting transition and then remaining at negative values (inset of Fig. 4(a)). Using STO as a back gate is expected to electrostatically modulate the superconductivity. Fig. 4(a) displays the resistance as a function of temperature at various gate voltages ($V_g$) for FeSe/STO(110). Under negative biases, the superconducting transition shifts steadily to lower temperatures and $T_{onset}$ decreases by 2.8 K at a voltage of -200 V. A much weaker enhancement of 0.5 K is present under positive gating, which may reflect that an optimal doping is nearly reached. Such a dichotomy was previously observed in the FeSe/STO(001) system[5], which is re-plotted in Fig. 4(b). Fig. 4(c) shows a comparison between FeSe/STO(110) and FeSe/STO(001), where the superconducting transition temperature $T_{onset}$ is plotted as a function of gate voltage $V_g$. Both substrates have a nominal thickness of 0.50 ±0.05 mm. The modulation of superconductivity with the electrostatic gating is smaller for FeSe/STO(110) interface than for FeSe/STO(001) interface, which will be further discussed.

The observation of high temperature superconductivity in FeSe/STO(110) provides us with some insights into the mechanism of the interface enhanced high temperature superconductivity. From the structure point of view, FeSe films on STO(110) exhibit a rectangular lattice, strikingly different from FeSe/STO(001) where ~3% expansion is observed for both directions[20]. Nevertheless, the superconducting gap observed here (17 meV) is comparable with the gap in FeSe/STO(001) within experimental uncertainty[2-4,30] or merely ~15% off the optimal value[1,31]. Strain can be therefore excluded from the critical factors for the interface high temperature superconductivity in FeSe/STO. The lattice variation would significantly affect the antiferromagnetic superexchange interactions between Fe moments[36]. Its minor role speaks against that the antiferromagnetic-interaction/spin-fluctuation is responsible for electron paring here. Rather, the high temperature superconductivity with comparable superconducting gaps in both FeSe/STO(110) and FeSe/STO(001) may be understood under the scenario of interface charge transfer and interface enhanced electron-phonon coupling as experimentally evidenced[2-4,6,16,17,19,20] and theoretically supported[6,15,18,21,37]. Both systems host similar 2D carrier densities (due to oxygen vacancies)[23,38], and a O-Ti-O stretching mode with the energy at ~ 100 meV[39,40], which couples with FeSe electrons and contributes to superconductivity as revealed by the ARPES study[6].

We have systematically studied the superconductivity properties of ten FeSe/STO(110) samples in total and found that FeSe/STO(110) always exhibits a slightly smaller superconducting gap and a lower transition temperature than FeSe/STO(001) grown in the same MBE systems does. This indicates a smaller interface enhancing factor for FeSe/STO(110) than for FeSe/STO(001). We speculate that the dielectric constant and the spatial extension of 2D carriers may play a role. On one hand, for the 2D superconductivity where the electrons are confined in the plane parallel to the FeSe/STO interface, in-plane dielectric constant $\varepsilon_\parallel$ reflects the strength of electron screening or Coulomb interactions. And according to the theoretical picture in Ref. *18*, the interface electron-phonon coupling constant is proportional to $\sqrt{\varepsilon_\parallel/\varepsilon_\perp}$, where $\varepsilon_\parallel$ and $\varepsilon_\perp$ are the dielectric constants parallel and perpendicular to the FeSe/STO interface, respectively. Our experimental results agree with the fact that $\varepsilon_\parallel$ ($\varepsilon_\perp$) of STO(110) is smaller (larger) than that of STO(001)[27,28]. On the other hand, the spatial extension of 2D carriers along the vertical direction is larger for STO(110) surface than for STO(001) surface[24]. This weaker confinement of 2D carriers should give rise to lower charge transfer to FeSe films and hence a lower $T_C$. It also explains the weaker modulation of superconductivity by electrostatic gating and the lower upper critical field in FeSe/STO(110).

In conclusion, we demonstrate the interface enhanced superconductivity in 1-UC FeSe films on STO(110) by *in situ* STS and direct transport study. FeSe/STO(110) exhibits similar high temperature superconductivity properties to the previously studied FeSe/STO(001) system. It suggests that interface charge transfer and interface enhanced electron-phonon coupling proposed in our previous study[1] and supported recently by several studies[2-4,6,16,18-21,37] may be responsible for the high temperature superconductivity observed in both FeSe/STO(110) and FeSe/STO(001) systems.

ACKNOWLEDGEMENT

This work is supported by NSFC (91421312, 11574174 and 11404183) and MOST of China (2015CB921000).

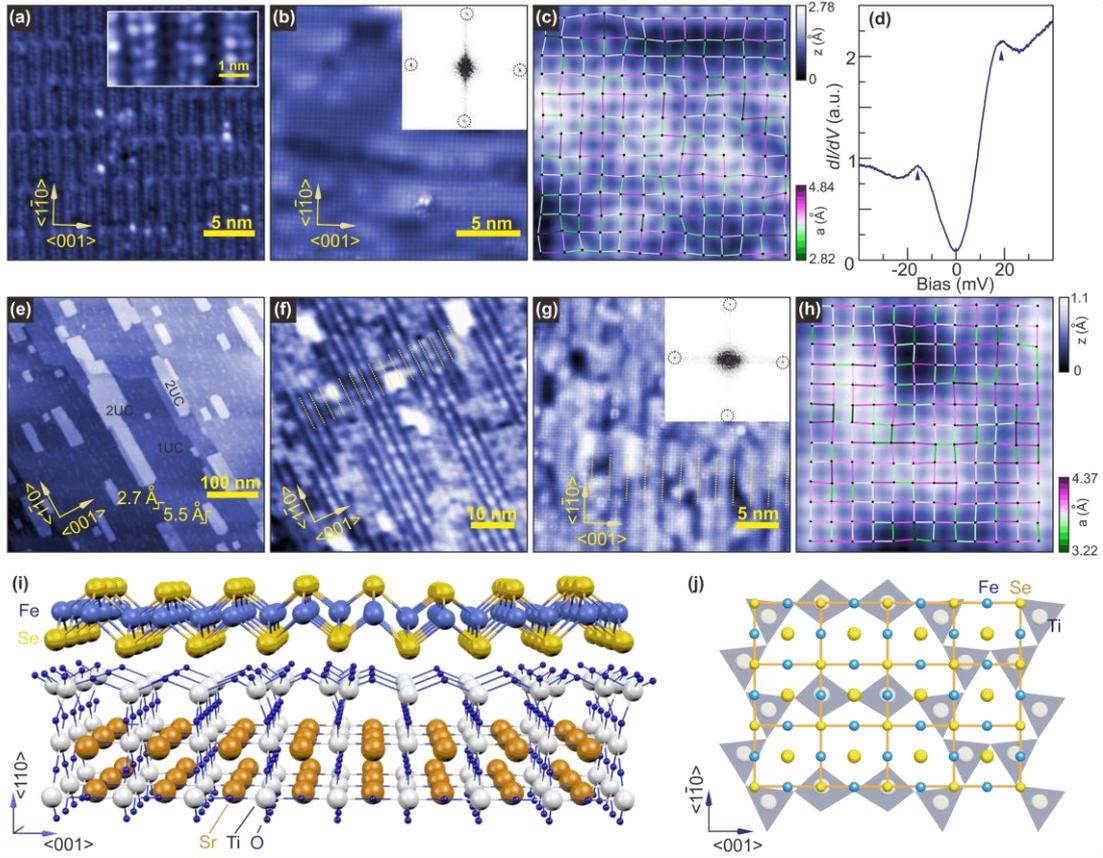

Fig. 1 (color online) (a) STM topography of Nb-STO-(4×1) surface ($V$=1 V, $I$=50 pA, inset: $V$=500 mV, $I$=100 pA)). (b)-(c) STM topography of 1-UC FeSe films on Nb-STO(110) substrate ((b)$V$=200 mV, $I$=100 pA and (c) $V$=30 mV, $I$=300 pA). (d) Typical dI/dV curves taken on FeSe films ($V$ = 30 mV, $I$ = 200 pA). The triangles show the positions of coherent peaks. (e)-(h) STM topography of 1-UC FeSe films on insulating STO(110) substrate ((e)$V$=1V, $I$=100 pA and (f) $V$=600 mV, $I$=100 pA on sample #1, (g) $V$=100 mV, $I$=50 pA and (h) $V$=300 mV, $I$=200 pA) on sample #2). The insets in (b) and (g) are the corresponding fast Fourier transform images (6.0 nm$^{-1}$ × 6.0 nm$^{-1}$). In (c) and (h), local maxima (black dots) are taken as the positions of Se atoms, the distance between adjacent atoms are manifested by the colored segments. Short lines in (f) and (g) are equidistant with a spacing of ~2.4 nm and ~2.0 nm, respectively, consistent with the period of (6×1) and (5×1) reconstructions of TiO$_4$ layer on SrTiO$^{4+}$ surface. Figures (a)-(d), (g) and (h) are acquired at 4.6 K and (e) and (f) at room temperature. (i) Sketch of 1-UC FeSe epitaxially grown on STO(110)-(4 × 1). (j) Schematic of the coincidence site lattice in 1-UC FeSe on STO(110)-(4 × 1).

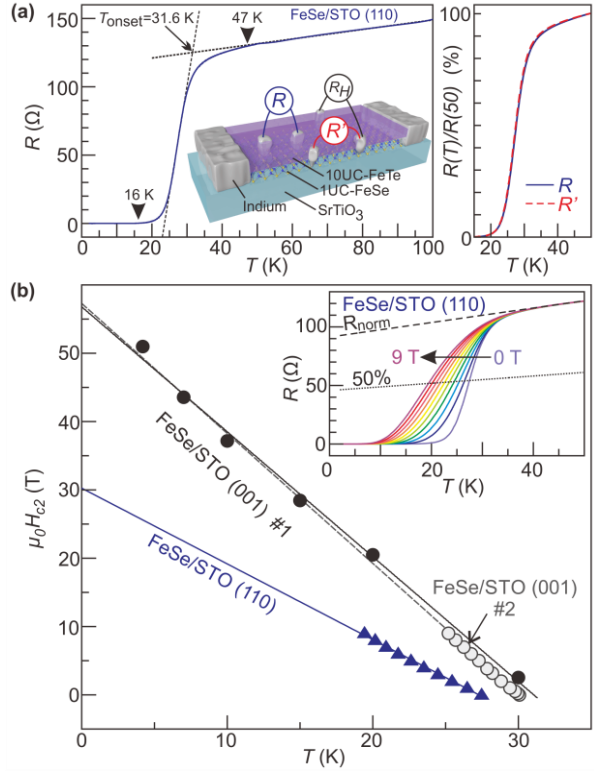

FIG. 2 (color online) (a) Resistance of the 10UC-FeTe/1UC-FeSe/STO(110) system as a function of temperature. Inset shows the schematic for measuring the longitudinal and the Hall resistance. Right panel shows that the resistances measured at two positions (center and edge as indicated in the sketch of the left panel) show exactly the same transition. (b) Upper critical field as a function of temperature for FeSe/STO(110) and FeSe/STO(001). Inset shows R(T) of FeSe/STO(110) at magnetic fields from 0 to 9 T with an increment of 1 T. $T_{mid}$ is defined as the temperature where the dotted line crosses with the R(T) curve. In the main panel, filled circles are obtained from a FeSe/STO(001) sample measured in a pulsed magnetic field, and empty circles in a static magnetic field. Straight lines are extrapolations of the linear fittings.

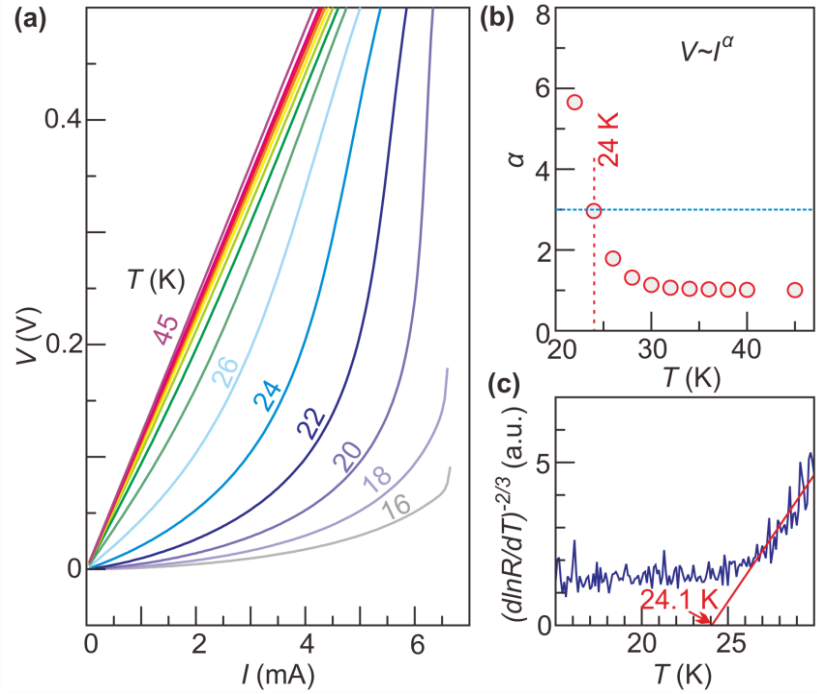

FIG.3 (color online) (a) Voltage as a function of current at different temperatures. Except for the data at 45 K, other curves are obtained from 40 to 16 K with a step of 2 K. (b) Temperature dependence of the power law exponent α determined from $V \sim I^\alpha$. The exponents are extracted from fittings to the data presented in (a). (c) R-T dependence on a $[d\ln(R)/dT]^{-2/3}$ scale. The Halperin-Nelson formula $R(T) = R\exp[-b(T/T_{BKT}-1)^{-1/2}]$ states that $[d\ln(R)/dT]^{-2/3}$ has a linear dependence on T above $T_{BKT}$. The red straight line is such a fit that extrapolates to zero at a BKT temperature of 24.1 K.

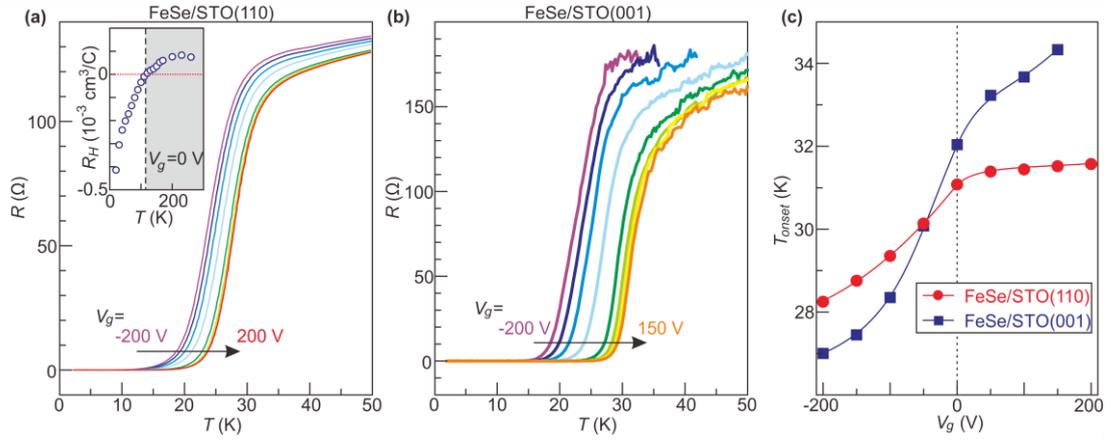

FIG. 4 (color online) (a) and (b) *R(T)* curves at various gate voltages measured on 10UC-FeTe/1UC-FeSe/STO(110)and a 10UC-FeTe/1UC-FeSe/STO(100) systems, respectively. Here both substrates have a nominal thickness of 0.5 ± 0.05 mm. Inset to (a) displays the Hall coefficient as a function of temperature for the FeSe/STO(110) system. Data in (b) is taken from ref. 16. (c) $T_{onset}$ as a function of gate voltage for the FeSe/STO(110) and FeSe/STO(001) systems.